\documentclass[seceq,supplement]{ptptex}

\title{
Analogies in theoretical physics%
}

\author{
Giovanni \textsc{Jona-Lasinio}%
}

\inst{Dipartimento di Fisica, ``Sapienza'', Universit\`a di Roma, \\
Piazzale A. Moro 2, Roma 00185, Italy\\
Istituto Nazionale di Fisica Nucleare, Sezione di Roma 1, \\
Roma 00185, Italy}

\abst{
Analogies have had and continue to have an important role in the 
development of theoretical physics. They may start from similarities
of physical concepts followed by similarities in the mathematical 
formalization or it may be a purely mathematical aspect to suggest
the development of analogous physical concepts. More often a subtle
non obvious interplay between these levels is involved.
In this paper I will discuss two cases sufficiently intricate to
illustrate some ways of how analogies work. The first topic is
the introduction of spontaneous symmetry breaking in particle physics.
The second one is the use of the renormalization group in the theory
of critical phenomena and its statistical interpretation.}

\begin{document}

\maketitle

\section{Introduction}

The transfer of ideas from one domain of science to another is a 
complicated process which involves that ill-defined concept that we 
call intuition. This is something very subjective and uses entirely 
different paths according to the cultural background and 
inclinations of each scientist. In theoretical physics an intuition 
may have as a starting point some imperfect or incomplete 
parallelism of physical concepts, but sometimes it is a 
mathematical analogy which is at the origin of a new development 
and physical concepts are shaped along the way. 
In this talk I will try to call your attention on the role played by 
analogies in the construction of some physical theories. 
Several cases can be found in the history of physics in which 
analogies were viewed as an important methodological tool.
It is interesting to recall Boltzmann's comments on Maxwell's theory: \cite{BO}

{\sl ``Most surprising and far-reaching analogies revealed themselves
between apparently quite disparate natural processes. It seemed that
nature had built the most various things on exactly the same pattern;
or, in the dry words of the analyst, the same differential equations 
hold for the most various phenomena. Thus thermal conduction, diffusion 
and the distribution of charge in electric conductors follow the
same laws. The same equations may be regarded as the solution of
a problem in hydrodynamics and in potential theory. The theory of
fluid vortices and that of friction in gases show the most surprising
analogy with electrodynamics an so on. (See also Maxwell, Scientific
Papers, Vol. I).

...........
In his very first paper on the 
theory of electricity (On Faraday's lines of force, see Scientific
Papers, Vol. I p. 157), Maxwell declares that he does not intend to propose
a theory of electricity; that is he does not himself believes in the
reality of the incompressible fluids and resistances that he is assuming,
but merely wishes to give a mechanical example that shows much analogy with
electric phenomena, which he wants to present in a form that makes them most 
readily
understandable. In his second paper (On physical lines of force, Scientific 
Papers, Vol. I p. 451) he goes much further still, constructing from fluid
vortices and and friction rollers moving inside cells with elastic walls an
admirable mechanism that serves as a mechanical model for electromagnetism.'' }

I wish to  recall also two very important more recent examples in which
mathematical analogy played a relevant role. In Dirac formulation
of quantum mechanics the Hamiltonian formalism reinterpreted in terms
of operator variables was a leading thread. The first quantum theories
of elementary particles, the theory of $\beta$-decay by Fermi and the theory
of nuclear forces by Yukawa, had as their model-guide the already established
QED.   

I will try to illustrate the role of analogy in contemporary theoretical physics
with the help of developments in which I have been involved. 
I hope that it may contribute to stimulate further reflection on these themes.
I will discuss in some detail the introduction
and formalization of the idea of spontaneous symmetry 
breaking (SSB) in
particle physics which is a case of cross fertilization
from condensed matter.  
As a second example I will
consider the use of the renormalization group (RG) in statistical mechanics
where the flow of ideas goes in the opposite
direction. The RG is a powerful tool for calculations and, suitably
reformulated in terms of probabilistic concepts, clarifies the
statistical meaning of  universality in critical phenomena.
This last development was again stimulated by an analogy, a mathematical one.   

\section{Spontaneous symmetry breaking in particle physics}
Spontaneous breakdown of symmetry is a concept that is 
applicable only to systems with infinitely many degrees of freedom. 
Although it pervaded the physics of condensed matter for a very 
long time, magnetism is a prominent example, its formalization 
and the recognition of its importance has been an achievement of 
the second half of the $XXth$ century. Strangely enough the name 
was adopted only after its introduction in particle physics: it is due 
to Baker and Glashow \cite{BG}. Very often concepts acquire a proper name 
only when they attain their full maturity. 

What is SSB? In condensed matter physics means that
the lowest energy state of a system can have a lower
symmetry than the forces acting among its constituents
and on the system as a whole. As an example consider
a long elastic bar on top of which we apply a compression force
directed along its axis. Clearly there is rotational symmetry
around the bar which is maintained as long as the
force is not too strong: there is simply a shortening
according to the Hooke's law. However when the force
reaches a critical value the bar bends
and we have an infinite number of equivalent
lowest energy states which differ by a rotation. 

Heisenberg was probably the first to consider
SSB as a possibly relevant concept in particle physics \cite{H}
but his proposal was not  the physically right one.
The theory of superconductivity
of Bardeen, Cooper and Schrieffer which
appeared in 1957 \cite{BCS} provided the key
paradigm for the introduction of SSB in
relativistic quantum field theory and particle physics on the basis of 
an analogy proposed by Nambu \cite{N1}. In his Nobel lecture Nambu \cite{N2}
emphasizes the importance of his previous exposure to condensed matter 
physics.

To appreciate the innovative character of this concept
in particle physics one should consider the strict
dogmas which constituted the foundation of relativistic quantum
field theory at the time. One of the dogmas stated that 
all the symmetries of the theory, implemented
by unitary operators, must leave the lowest energy state, the vacuum, invariant.
This property does not hold in presence of SSB 
and degenerate vacua. These vacua cannot be connected by local
operations and are orthogonal to each other giving rise to different
Hilbert spaces. If we live in one of them 
SSB will  be manifested by
its consequences, in particular the particle spectrum.

\medskip

The BCS theory of superconductivity, immediately after its appearance, 
was reformulated and developed by 
several authors including Bogolyubov,
Valatin, Anderson, Ricayzen and Nambu.
The following facts were emphasized

\medskip

\begin{itemize}
\item[1.]
The ground state proposed by BCS is not invariant under gauge
transformations.
\item[2.]
The elementary fermionic excitations (quasi-particles) are not 
eigenstates of the charge as they appear as a superposition
of an electron and a hole.
\item[3.]
In order to restore charge conservation these excitations must be
the source of bosonic excitations described by a long range  (zero mass) 
field. In this way the original gauge invariance of the theory is
restored.
\end{itemize}

\medskip

The peculiarity of the paper of 
Nambu \cite{N0}, was that he
used a language akin to quantum field theory, that is the 
Green's functions formalism, and the role of gauge invariance was
discussed in terms of vertex functions and the associated Ward identities. 
The search for
analogies in particle physics became quite natural. In particular, following
the suggestion of \cite{N1},
the study of  chiral symmetry breaking   was  
developed in  detail in two papers by Nambu and Jona-Lasinio 
\cite{NJL1,NJL2} which had a considerable influence on the evolution
of elementary particle theories.

Let us illustrate the elements of the analogy.

\medskip

 Electrons near the Fermi surface are described by the following equation
  \begin{eqnarray}
\begin{split}
    E \psi_{p,+} &=& \epsilon_p \psi_{p,+} + \phi \psi_{-p,-}^\dag
    \\
    E \psi_{-p,-}^\dag &=& -\epsilon_p \psi_{-p,-}^\dag + 
    \phi \psi_{p,+},
\end{split}
  \end{eqnarray}
  with eigenvalues
  \begin{eqnarray}
    E = \pm\sqrt{\epsilon_p^2+\phi^2}.
  \end{eqnarray}
  Here, 
  $\psi_{p,+}$ and $\psi_{-p,-}^\dag$ are the wavefunctions for an electron 
  and a hole of momentum $p$ and spin $+$; $\phi$ is the gap.

\medskip

In the Weyl representation, the Dirac equation reads
  \begin{eqnarray}
\begin{split}
    E \psi_{1} &=& \pmb{\sigma}\cdot\pmb{p} \psi_{1} + m \psi_{2}
    \\
    E \psi_{2} &=& -\pmb{\sigma}\cdot\pmb{p} \psi_{2} + m \psi_{1},
\end{split}
  \end{eqnarray}

with eigenvalues
  \begin{eqnarray}
    E = \pm\sqrt{p^2+m^2}.
  \end{eqnarray}
  Here, $\psi_1$ and $\psi_2$ are the eigenstates of the chirality operator 
  $\gamma_5$. Particles
with mass are superpositions of states of opposite chirality. The similarity
is obvious.

\medskip

The bosonic excitations necessary to restore gauge invariance in a 
superconductor appear in the 
approximate expressions for the charge density and the current
in a BCS superconductor \cite{N0}, 
  \begin{eqnarray}
\begin{split}
    \rho(x,t) &\simeq& \rho_0 + 
    \frac{1}{\alpha^2} \partial_t f
    \\
    \pmb{j}(x,t) &\simeq& \pmb{j}_0 - \pmb{\nabla} f ,
\end{split}
  \end{eqnarray}
where $\rho_0=e \Psi^\dag \sigma_3 Z \Psi$ and 
  $\pmb{j}_0=e \Psi^\dag (\pmb{p}/m) Y \Psi$ are the contributions
of the quasi-particles, $Y$, $Z$, $\alpha$ are
constants and $f$ satisfies the wave equation
  \begin{eqnarray}
    \left( \nabla^2 -\frac{1}{\alpha^2} {\partial_t}^2 \right) f
    \simeq -2 e \Psi^\dag \sigma_2 \phi \Psi.
  \end{eqnarray}
Here, $\Psi^\dag=(\psi^\dag_1,\psi_2)$

\medskip

In the elementary particle context
the axial current  $\bar\psi \gamma_5 \gamma_\mu \psi$ 
is the analog of the electromagnetic current
in BCS theory. In the hypothesis of exact conservation, the matrix elements 
of the axial current between nucleon states of four-momentum $p$ and $p'$ 
have the form
  \begin{eqnarray}
    \varGamma_\mu^A(p',p)= \left( i \gamma_5 \gamma_\mu - 
      2m \gamma_5 q_\mu / q^2 \right)F(q^2),
    \qquad q=p'-p .
  \end{eqnarray}
Exact conservation is compatible with a finite nucleon mass $m$
provided there exists a massless pseudoscalar particle. 

Assuming exact conservation of the chiral current,
a picture of chiral SSB may consist in a vacuum of
a massless Dirac field viewed as a sea of occupied negative
energy states, and an attractive force between particles and
antiparticles  having the effect of
producing a finite  mass, the counterpart
of the gap. The pseudoscalar massless particle, which may be interpreted
as a forerunner of the pion, corresponds to the  bosonic field associated to the
fermionic quasi-particles in a superconductor.  

To  implement this picture the construction of a relativistic
field theoretic model was required.
At that time Heisenberg and his
collaborators had developed a comprehensive theory
of elementary particles based on a non linear spinor
interaction: the physical principle was  that spin $\frac {1}{2}$
fermions could provide the building blocks of all
known elementary particles. Heisenberg was however
very ambitious and wanted at the same time to
solve in a consistent way the dynamical problem
of a non renormalizable theory. This made their approach
very complicated and not transparent. Nambu considered Heisenberg theory
very formal, but the four spinor interaction was
attractive due to its simplicity and analogy with
the many-body case. I was more enthusiast.
I had been exposed
several times to the nonlinear spinor theory, first in
a meeting in Venice where a very interesting discussion
between Heisenberg and Pauli took place, then in Rome
that Heisenberg visited just to explain his theory. At that time
I believed in such fundamental theories!

\medskip

A Heisenberg type Lagrangian was adopted without
pretending to solve the non-renormalizability problem
and introducing a relativistic cut-off to cure the
divergences. This model is known in the literature with the acronym NJL.
The energy scale of interest was of
the order of the nucleon mass and one hoped that
higher energy effects would not change substantially
the picture.

\medskip

The Lagrangian of the NJL model is
  \begin{eqnarray}
    {\mathcal L} = -\bar\psi \gamma_\mu \partial_\mu \psi+
    g\left[ (\bar\psi\psi)^2-(\bar\psi\gamma_5\psi)^2\right].
  \end{eqnarray}
It is invariant under ordinary and chiral gauge transformations
  \begin{eqnarray}
\begin{split}
    &&\psi \to e^{i\alpha} \psi, 
    \qquad
    \bar\psi \to \bar\psi e^{-i\alpha}
    \\
    &&\psi \to e^{i\alpha\gamma_5} \psi, 
    \qquad
    \bar\psi \to \bar\psi e^{i\alpha\gamma_5}.
\end{split}  
\end{eqnarray}

\medskip

To investigate the content of the model a simple mean field
approximation for the mass was adopted
\begin{eqnarray}
\begin{split}
&&m=2g[<\bar \psi \psi> - \gamma_5 <\bar \psi \gamma_5 \psi>]\\
&&=-2g[tr S^{(m)}(0) - tr \gamma_5  S^{(m)}(0)],
\end{split}
\end{eqnarray}
where $S^{(m)}$ is the propagator of the Dirac field of mass $m$,
or more explicitly,
\begin{eqnarray}
\frac{2\pi^2}{g\Lambda^2}= 1-\frac{m^2}{\Lambda^2}
\ln\left(1+\frac{\Lambda^2}{m^2}\right),
\end{eqnarray}
where $\Lambda$ is the invariant cut-off. This equation is very
similar to the gap equation in BCS theory. If $\frac{2\pi^2}{g\Lambda^2} < 1$
there exists a solution $m>0$.
  
\medskip
 
From this relationship a rich spectrum of bound states follows,

\bigskip

    \begin{center}
      \begin{tabular}{cccc}
        \hline\hline
        nucleon & mass $\mu$ & spin-parity & spectroscopic \\
        number & & & notation \\
        \hline
        $0$ & $0$ & $0^-$ & $^1S_0$\\
        $0$ & $2m$ & $0^+$ & $^3P_0$\\
        $0$ & $\mu^2 > \frac83 m^2$ & $1^-$ & $^3P_1$\\
        $\pm 2$ & $\mu^2>2 m^2$ & $0^+$ & $^1S_0$\\
        \hline\hline
      \end{tabular}
    \end{center}
 
\bigskip

The bosonic field in the superconductor and the pseudoscalar particle in the
NJL model are special cases of a general theorem formulated by Goldstone 
in 1961 \cite{G}.

\medskip

\emph{Whenever the original Lagrangian has a continuous symmetry group,
which does not leave the ground state invariant,
massless bosons appear in the spectrum of the theory.} 

\medskip

Other examples are,

\medskip
 
    \begin{center}
      \begin{tabular}{lll}
        \emph{physical system} & \emph{broken symmetry} & \emph{massless bosons}\\
        ferromagnets & rotational invariance & spin waves\\
        crystals & translational and rotational invariance & phonons\\
      \end{tabular}
    \end{center}

\medskip

These massless bosons are now known in the literature as Nambu-Goldstone bosons.
In nature, however, the axial current is only approximately conserved.
The model could make contact with the real world under the hypothesis  
that the small violation of axial current conservation
gives a mass to the massless boson, which is then identified with the $\pi$ 
meson.

\medskip

Later, a reinterpretation in terms of quarks of the NJL model provided  
a successful effective theory of low energy Quantum Chromodynamics, 
see for example \cite{HK}. 
After the NJL model, SSB  became a key ingredient 
in elementary particle physics. Electroweak unification \cite{W}
is based on a mechanism which has also
its roots in the theory of superconductivity. For this mechanism we refer 
to the papers by Anderson,
Brout, Englert and Higgs \cite{A,BE,HI}. This important part of the story  
however goes beyond the purpose of the present talk.

\subsection{\bf The effective action}
The argument showing that SSB actually takes place in the NJL model was 
based on a self-consistent field approximation and a formulation independent
of any kind of approximation was desirable. This was the first motivation
of my paper \cite{JL}.

\medskip

\medskip

In field theory we can define a formal analog of the partition function
\begin{eqnarray}
Z(J)=\langle 0|T e^{i(\int dx ({\mathcal L_I} + \sum J_i \Phi_i))}|0 \rangle,
\end{eqnarray}
where $|0\rangle$ is the bare vacuum, the $J_i$ are external sources and 
the fields $\Phi_i$ transform according to a representation of some symmetry 
group, e.g., the fundamental 
representation of the orthogonal group. Its logarithm,
\begin{eqnarray}
G(J)=-i\log Z(J),
\end{eqnarray}
is the generator of the connected time ordered Green's functions 
  (in statistical mechanics the analog of $G$ is the free energy in presence 
  of an external field $J$). Define the \emph{``classical''} fields  
  \begin{eqnarray}
    \frac{\delta G}{\delta J_i} = \langle \Phi_i \rangle = 
    \phi_i.
  \end{eqnarray}
Assuming that this relationship can be inverted,
the effective action is the dual functional $\Gamma[\phi]$ 
defined by the Legendre transformation
\begin{eqnarray}
\Gamma (\phi) = G(J(\phi))- \int dx \sum J_i(\phi) \phi_i.
\end{eqnarray}
We have the conjugate relation to (2.14)
  \begin{eqnarray}
    \frac{\delta \Gamma}{\delta \phi_i} = -J_i.
  \end{eqnarray}
The vacuum of the theory is defined by the variational equation
  \begin{eqnarray}
    \frac{\delta \Gamma}{\delta \phi_i} = 0.
  \end{eqnarray}
$\Gamma[\phi]$ is the generator of the one particle irreducible vertex 
functions and can be constructed by simple diagrammatic rules.

A symmetry breaking solution is a solution $\phi_i$ of the variational equation 
transforming in a non trivial way under the symmetry group.
This is the way in which SSB is nowadays introduced in textbooks
\cite{IZ,PS,MIR,ZJ}.
The effective action provides the natural setting to analyse the stability 
of the solutions of the variational problem (2.9). Furthermore 
the derivation of the Goldstone theorem in this formalism is particularly
simple \cite{JL}. 

\medskip

Now some history. The similarity of the formalisms
of quantum field theory and statistical mechanics was part of the common
wisdom, it had been emphasized for instance in the book by Bogolyubov
and Shirkov \cite{BOS}.  
A characteristic feature of statistical mechanics,
both classical and quantum, is the existence of
variational principles determining the stable
states of a system. Variational principles
in quantum statistical mechanics have been introduced
by Lee and Yang \cite{LY} followed by Balian, Bloch and De Dominicis 
\cite{BBD} and by De Dominicis and Martin \cite{DM}. 
The variables appearing in these principles are
typically quantum averages of operators, that is c-numbers.
The work of De Dominicis and Martin used functional methods
typical of Schwinger's school with which I had some
familiarity.  
A general tool to derive such variational principles  
was the functional Legendre transform with respect
to space-time dependent potentials. 
I found then natural to introduce the effective
action, a c-number action functional for quantum
field theory, 
to characterize the vacuum in terms of a variational principle. 
For homogeneous systems
the more restricted concept of effective potential, a limiting case of the 
effective action, had been introduced via perturbation theory
in \cite{G} with a more systematic treatment in \cite{GSW}. 
However the effective potential does not provide a complete
description of the dynamics. The effective action
differs from a classical action as it is non local in time and involves the
whole history of the system. An interpretation of the dynamical
equations in terms of an initial value problem is possible
only in certain limiting cases \cite{CJLPT}. 

Many years  after my first paper  I learnt that the effective action 
in a perturbative form
had appeared in a work by Heisenberg and Euler in the thirties \cite{HE}
where they studied quantum corrections
to the Maxwell equations. I also heard from Bryce DeWitt that it was 
considered by Schwinger in unpublished notes. However the usefulness
of the effective action was fully appreciated only after its introduction
in connection with SSB.
Inspired by the work of De Dominicis and Martin I studied also 
higher order Legendre transforms,
i.e. involving  expectation values of composite operators,
and the associated variational principles in quantum
field theory \cite{DaJL}. Their study was pursued later in a work
by Cornwall, Jackiw and Tomboulis \cite{CJT}.

\medskip

\section{Critical phenomena and renormalization group}
Statistical mechanics describes macroscopic systems in terms of an
underlying microscopic structure whose configurations are the
arguments of a probability distribution, an ensemble in the
terminology of physicists. 
When a system approaches a critical point large islands
of a new phase appear so that correlations among the microscopic
constituents extend over macroscopic distances. One characterizes
these situations by introducing a correlation length which measures
the extension of such correlations. At the critical point
this length becomes infinite and typically correlations decay
with a non-integrable power law as opposed to an exponential
decrease away from criticality. 

The exponents in these power laws
exhibit a remarkable degree of universality because to systems
physically different such as a gas and a ferromagnet correspond the
same exponents, e.g. the magnetization,
\begin{eqnarray}
\nonumber
m=A_m |T-T_c|^{\beta},
\end{eqnarray}
and the difference between the liquid and the gas densities,
\begin{eqnarray}
\nonumber
\rho_L - \rho_G = A_{LG} |T - T_c|^{\beta},
\end{eqnarray}
are characterized by the same power law.

In this case there was a
transfer of ideas from quantum field theory to
many-body theory and statistical mechanics. 
In 1966 there was an important school on
critical phenomena at Brandeis University 
where Kadanoff explained his ideas on the
origin of critical scaling and the ensuing equations
for the structure of the correlation functions.
This school was attended by Carlo Di Castro, a
young many-body physicist at that time, and when
he came back he told me about Kadanoff theory.
My reaction was that Kadanoff's scaling equations for 
the correlation functions
looked like a simplified version of the multiplicative
renormalization group (RG) equations satisfied by Green's functions in
quantum field theory and  statistical mechanics. 
I thought that  mathematically, scaling and universality should arise
from a resummation of singularities similarly
to what happens in certain field theoretic infrared problems. This provided
the basis for an analogy.  

Kadanoff's qualitative argument to explain why scaling properties
should   be expected at the critical point was the following: 
if correlations extend over macroscopic
distances it must be irrelevant whether we consider our system
constituted by the original microscopic objects or by
blocks containing a large number of constituents \cite{KA}. In the limit
when the correlations extend to infinity the size of the blocks
should not matter and this leads  to homogeneity
properties for the correlation functions  and other 
thermodynamic quantities. 

The equation of the multiplicative renormalization group used  
in quantum field theory in the simplest case has the following form \cite{BOS}
\begin{equation}
d(x,y,\alpha)= Z(t,y,\alpha) ~
d(x/t,y/t,\alpha Z_V^{-1}(t,y,\alpha)Z^2(t,y,\alpha)),
\label{DJ}
\end{equation}
where $d(x,y,\alpha)$ is a dimensionless two-point Green function depending
on a momentum squared $x=p^2$,  a mass parameter $y=m^2$ and
the intensity of the interaction $\alpha$ (dimensionless). 
The scaling functions
$Z$ and $Z_V$ can be expressed in terms of the  Green's functions themselves
via  normalization conditions at $p^2=t$ for $d(x,y,\alpha)$ and the dressed 
interaction (vertex function).
This is an
exact generalized scaling relation  which has a counterpart for the correlation
functions in statistical 
mechanics \cite{BBT}. Di Castro and I expected that in the vicinity of the
critical point this relationship would reduce to the 
phenomenological scaling 
due to the irrelevance of the coupling constant and the other  parameters. 
The first use of this equation
in the study of critical phenomena appeared in our 1969 paper \cite{DJL} and
we obtained in this way a qualitative explanation and foundation of scaling
from first principles. 
After the introduction of a non integer space dimension $d$ and 
the introduction of $\epsilon=4-d$ as a perturbation parameter \cite{WF},
the multiplicative RG became
the basis for systematic quantitative calculations \cite{DI,MI,BR}.

About two years after our paper, an article by Wilson \cite{WI} appeared 
where a  notion
of renormalization group apparently different was used. Actually this 
notion had been used before by Wilson in connection with the fixed source
meson theory.   
This notion looked  closer than the multiplicative RG
to Kadanoff's picture described above. However the mathematically most faithful 
implementation of Kadanoff's idea came from another direction and again
an analogy provided the key idea.

\subsection{\bf Renormalization group and probability theory}
Forming blocks of stochastic variables, as in Kadanoff's picture,
is common practice
in probability, the central limit theorem (CLT) being the prototype
of such a way of reasoning.  
CLT asserts the following. Let $\xi_1, \xi_2,\ldots,\xi_n,\ldots$ be a
sequence of independent identically distributed (i.i.d.) random
variables with finite variance $\sigma^2={\mathbb{E}}
(\xi_i - ({\mathbb{E}}(\xi_i))^2$,
where ${\mathbb{E}}$ means expectation with respect to their common distribution. Then
\begin{eqnarray}
{{\sum_1^n (\xi_i - {\mathbb{E}}(\xi_i))}\over{\sigma n^{1/2}}} 
\stackrel{n \to \infty}{\longrightarrow} N(0,1),
\end{eqnarray}
where the convergence is in law and $N(0,1)$ is the normal centered
distribution of variance $1$.

The crucial point is that when 
we sum many random variables we have to normalize properly
the sum in order to obtain a regular probability distribution.
In the case of the CLT the correct normalization is proportional to
the square root of the number of variables, and represents the square root
of the variance of the sum. 
When we deal with processes which have  correlations
the variance can be written
\begin{eqnarray}
{\mathbb{E}}((\sum_i \xi_i )^2 )=N\sigma^2 + N\sum_j {\mathbb{E}}(\xi_0 \xi_j),
\end{eqnarray}
where we have assumed translational invariance. The sum in the
second term is the susceptibility which diverges at the critical point
and dominates over the first term. We must therefore change the
normalization. The normalization is directly related to the rescaling
of the variables in the RG.

We now describe a RG derivation \cite{JL3} of the CLT and then explain how
an analogy and a generalization provide an interpretation of
the Kadanoff-Wilson view which unveils the statistical meaning
of universality of critical exponents and at the same time the
connection with  equation (3.1). 

\subsection{\bf {A renormalization group derivation of the central limit theorem}}

To visualize things consider the random variables $\xi_i$ as
discrete or continuous spins associated to the points of a
one dimensional lattice $\bf {Z}$ and introduce the block variables
$\zeta_n^1 = 2^{-{n/2}} \sum_1^{2^n} \xi_i$ and $\zeta_n^2 = 2^{-{n/2}}
\sum_{2^n+1}^{2^{n+1}} \xi_i$ 
Then
\begin{eqnarray}
\zeta_{n+1} = {\frac{1} {\sqrt{2}}}  (\zeta_n^1 +  \zeta_n^2).
\end{eqnarray}
Therefore we can write the recursive relation for the corresponding
distributions
\begin{eqnarray}
p_{n+1}(x)=\sqrt{2}\int dy~ p_n(\sqrt{2}x-y)p_n(y) = ({\cal R} p_n)(x).
\label{R}
\end{eqnarray}
The non linear transformation ${\cal R}$ is what we call a renormalization
transformation.

 Let us find its fixed points, i.e. the solutions of the
equation ${\cal R} p=p$. An easy calculation shows that the family
of Gaussians
\begin{eqnarray}
p_{G,\sigma}(x)={{1}\over {\sqrt{2\pi \sigma^2}}} e^{-{{x^2}\over
{2\sigma^2}}},
\end{eqnarray}
are fixed points. 
To prove the CLT we have to discuss the
conditions under which the iteration of ${\cal R}$ converges to a fixed 
point of variance $\sigma^2$. This amounts to determining the 
so-called \emph{domain of attraction} of the normal law.

The standard analytical way to prove the CLT is the Fourier  transform. 
Here
we shall illustrate the mechanism of convergence in the neighborhood of
a fixed point from the point of view
of nonlinear analysis. There are three conservation laws associated
with ${\cal R}$: normalization, centering and variance. In formulas
\begin{eqnarray}
\begin{split}
&&\int p_{n+1}(x)dx = \int p_{n}(x)dx\\
&&\int xp_{n+1}(x)dx = \int xp_{n}(x)dx\\ 
&&\int x^2p_{n+1}(x)dx = \int x^2p_{n}(x)dx .
\end{split}
\end{eqnarray}
Therefore only distributions with variance $\sigma^2$ 
can converge
to a Gaussian $p_{G,\sigma}(x)$. 
We fix $\sigma=1$ and write $p_{G}$ for $p_{G,1}$.

Let us write the
initial distribution as a centered deformation of the Gaussian
with the same variance
\begin{eqnarray}
p_{\eta}(x)=p_{G}(x)(1+\eta h(x)),
\end{eqnarray}
where $\eta$ is a parameter. The function $h(x)$ must satisfy
\begin{eqnarray}
\begin{split}
&&\int p_G(x)h(x)dx=0\\
&&\int p_G(x)xh(x)dx=0\\
&&\int p_G(x)x^2h(x)dx=0.
\end{split}
\end{eqnarray}

Suppose now  $\eta$  small. In linear approximation we have
\begin{eqnarray}
({\cal R} p_{\eta})= p_G (1+\eta({\cal L} h)) + {\cal O}(\eta^2),
\end{eqnarray}
where $\cal L$ is the linear operator
\begin{eqnarray}
({\cal L} h)(x)=2 \pi^{-1/2} \int dye^{-y^2}h(y+x2^{-1/2}).
\end{eqnarray}
The eigenvalues  of $\cal L$ are
\begin{eqnarray}
\lambda_k = 2^{1-k/2},
\end{eqnarray}
and the eigenfunctions the Hermite polynomials. The three
conditions above on $h(x)$ can be read as the vanishing of its
projections on the first three Hermite polynomials.

The mechanism of convergence of the deformed distribution to
the normal law is now clear in linear approximation: if we develop
$h$ in Hermite polynomials only terms with $k > 2$ will appear so that
upon iteration of the RG transformation they will contract
to zero exponentially as the corresponding eigenvalues are $< 1$.

The Gaussian  belongs to a very important class of distributions
called \emph{stable distributions}. They are characterized by
the fixed point equation
\begin{equation}
p(ax+b)={\frac {a_1 a_2}{a}} \int dy p(a_1(x-y)+b_1)p(a_2 y + b_2),
\end{equation}
where $a,a_1,a_2$ are positive numbers.
In the next subsection we shall briefly describe how to introduce
an analogous concept in the case of strongly dependent variables
as those appearing at the critical point in phase transitions. 
The concept of \emph{stable or self-similar random field} will
correspond to that of stable distribution.

\subsection{\bf{Strongly dependent variables: interacting spins at the critical point}}

The notion of self similar random field of discrete argument was introduced 
informally in \cite{JL1} to provide a proper mathematical setting 
for the notion of RG {\it a la} Kadanoff-Wilson. In rigorous form it was
described in \cite{JLG} followed by
\cite{SI}. In the present exposition we follow \cite{DO,SI1}. 

Let ${\bf {Z}}^d$ be a lattice in $d$-dimensional space and $j$ a generic
point of ${\bf {Z}}^d$, $j=(j_1, j_2, ...,j_d)$ with integer coordinates
$j_i$. We associate to each site a centered random variable $\xi_j$ and
define
a new random field, \emph{block-spin}
\begin{equation}
\xi_j^n=({\cal R}_{\alpha,n}\xi)_j=n^{-d\alpha/2}\sum_{s\in V_j^n} \xi_s,
\label{RF}
\end{equation}
where
\begin{equation}
V_j^n=\{s: j_kn - n/2 < s_k \leq j_kn + n/2 \},
\end{equation}
and $1\leq\alpha <2$. The transformation (\ref{RF}) induces a
transformation on probability measures according to
\begin {equation}
({\cal R}^*_{\alpha,n}P)(A)={P'}(A)=P({\cal R}^{-1}_{\alpha,n}A),
\label{RMU}
\end{equation}
where $A$ is a measurable set and ${\cal R}^*_{\alpha,n}$ has the semigroup 
property
\begin{equation}
{\cal R}^*_{\alpha,n_1}{\cal R}^*_{\alpha,n_2}={\cal R}^*_{\alpha,n_1n_2}.
\end{equation}
A measure $P$ will be called  self-similar if
\begin{equation}
{\cal R}^*_{\alpha,n}P=P,
\label{ST}
\end{equation}
and the corresponding field will be called a  self-similar random
field.
We briefly discuss the choice of the parameter $\alpha$.
It is natural to take $1\leq\alpha<2$. In fact $\alpha=2$ corresponds
to the law of large numbers so that  
the block variable (\ref{RF}) will tend for large $n$ to zero in probability.
The case $\alpha >1$
means that we are considering random systems which fluctuate 
more than a collection of independent variables and $\alpha=1$ 
corresponds to the CLT. Mathematically the lower bound is not
natural but it becomes so when we restrict ourselves to the 
consideration of ferromagnetic-like systems.

A theory of self similar random fields of generality comparable to the case
of stable distributions so far does not exist
and presumably is very difficult. However Gaussian fields are
completely specified by their correlation function and self similar
Gaussian fields can be constructed explicitly \cite{SI1}.

The search of non-Gaussian self-similar fields is considerably more difficult. A reasonable question is whether such fields exist in the
neighborhood of a Gaussian one. The approach to this problem
developed by physicists, the so called $\epsilon$ expansion, 
in our context can be interpreted as follows. 

Consider a
deformation $P_G(1+h)$ of a Gaussian self-similar field $P_G$ and the action of ${\cal R}^*_{\alpha,n}$
on this distribution. It is easily seen that
\begin{equation}
{\cal R}^*_{\alpha,n} P_G h =  {\mathbb{E}}(h|\{\xi_j^n\}) {\cal R}^*_{\alpha,n} P_G =
{\mathbb{E}}(h|\{\xi_j^n\}) P_G(\{\xi_j^n\}).
\label{LR}
\end{equation}
The conditional expectation on the right hand side of (\ref{LR})
will be called the linearization of the RG at $P_G$ and we want to study
its stability as a linear operator. For this purpose we have to
find the eigenvectors and eigenvalues of ${\mathbb{E}}(h|\{\xi_j^n\}$ . These have been calculated
by Sinai. The eigenvectors are appropriate infinite dimensional
generalizations of Hermite polynomials $H_k$ which are described in full
detail in \cite{SI1}. They satisfy the eigenvalue equation
\begin{equation}
{\mathbb{E}}(H_k|\{\xi_j^n\}) = n^{[k(\alpha /2 - 1) + 1]d}H_k(\{\xi_j^n\}).
\label{FPE}
\end{equation} 
We see immediately that  $H_2$ is always unstable, it is a relevant direction
in the physicist terminology.  The direction $H_4$ becomes
unstable when $\alpha$ crosses from below the value $3/2$. 
Bifurcation theory tells us that generically we must expect an exchange 
of stability between two fixed points and we should look for the new one
in the direction which has just become unstable.
By introducing
the parameter $\epsilon = \alpha - 3/2$, one can construct a non Gaussian fixed point using $\epsilon$ as a perturbation parameter. 
The formal construction is explained in
Sinai's book \cite{SI1} where one can find also a discussion of the questions, 
mostly still unsolved, arising in this connection.

\subsection{\bf Universality}
The previous analysis shows that in the probabilistic interpretation of the RG
universality of critical phenomena acquires a  clear statistical 
interpretation. In analogy with the case of the CLT there will be
different Gibbs distributions that under the RG will converge to the same
limit ensemble. A physical universality class will correspond to
a subset of the domain of attraction of the limit ensemble. 
In general we expect to be
a subset because not all distributions in the domain of attraction
will admit a natural physical interpretation.

\subsection{\bf {Multiplicative Structure}}
In this section we show that there is a natural multiplicative
structure, in mathematics called a cocycle, associated with transformations 
on probability
distributions induced by the block transformation. 
This multiplicative
structure is related to the properties of conditional expectations
\cite{CJL,JL3}.
Suppose we wish to evaluate the conditional expectation
\begin{equation}
{\mathbb{E}}(h|\{\xi_j^n\}),
\end{equation}
where the collection of block variables $\xi_j^n$ indexed by $j$
is given a fixed value. Here $h$ is a function of the  spins
$\xi_i$. It is an elementary property of conditional expectations 
that
\begin{equation}
{\mathbb{E}}({\mathbb{E}}(h|\{\xi_j^n\})|\{\xi_j^{nm}\}) = {\mathbb{E}}(h|\{\xi_j^{nm}\}).
\label{EX}
\end{equation}
Let $P$ be the probability distribution of the $\xi_i$ 
and ${\cal R}^*_{\alpha,n}P$ the distribution obtained by applying the RG transformation,
that is the distribution of the block variables $\xi^n_j$. By
specifying in (\ref{EX}) the distribution with respect to which
expectations are taken we can rewrite it as
\begin{equation}
{\mathbb{E}}_{{\cal R}^*_{\alpha,n}P}({\mathbb{E}}_P(h|\{\xi_j^n\})|\{\xi_j^{nm}\}) =
{\mathbb{E}}_P(h|\{\xi_j^{nm}\}).
\label{EX1}
\end{equation}
This is the basic equation of this section and we want to work out
its consequences. 
In analogy with the theory of
dynamical systems we interpret the conditional expectation as
a linear transformation  from the linear space tangent to $P$ to
the linear space tangent to ${\cal R}^*_{\alpha,n}P$ and we assume that
in each of these spaces there is a basis of vectors $H^{P}_k$,
$H^{{\cal R}^*_{\alpha,n}P}_k$  connected by
the following generalized eigenvalue equation \cite{OS}
\begin{equation}
{\mathbb{E}}_P(H^P_k|\{\xi_j^n\}) = \lambda_k(n, P) H^{{\cal R}^*_{\alpha,n}P}_k(\{\xi_j^n\}).
\label{GEE}
\end{equation}
Equation (\ref{EX1}) implies that the $\lambda$'s must satisfy
the relationship
\begin{equation}
\lambda_k(m, {\cal R}^*_{\alpha,n} P)\lambda_k(n, P) = \lambda_k(mn, P).
\label{J}
\end{equation}   
If $P$ is self-similar (\ref{J}) implies that the $\lambda$'s
are powers of $n$. An example is provided by (\ref{FPE}).
In the theory of the critical point the corresponding
eigenvectors are called scaling fields. When $P$ is not selfsimilar
the $\lambda$'s can be expressed in terms of suitable 
correlation functions.

The multiplicative
renormalization group of quantum field theory and statistical
mechanics, equation (3.1), is structurally similar to (\ref{J}).
It corresponds to a
simple transformation of the probability distribution leaving 
its form  unchanged while the values of its parameters 
are rescaled together with the random variables. 

\section{Conclusion}

I will conclude with a remark of a more general character.
New ideas  are not always immediately understood and 
I would like to  point out an
aspect relevant in their evolution: 
this is the language in which an idea is proposed.
The spontaneous
breakdown of a symmetry was rapidly absorbed by particle physicists:
the NJL model was formulated in
the standard language of the particle physics community, quantum field theory. 

The situation was somewhat different with the renormalization group. 
The RG for the first time provided a microscopic theory
of critical phenomena of wide applicability but required on the part of the 
interested 
community to adapt  to a new way of looking at the problems and 
to a new language originated in particle physics. 
This was a collective effort and happened in a remarkably 
short time. 

An important consequence in both cases was that condensed matter
and particle physicists became closer in their way of thinking.

\medskip

Coming back to analogies, they have  a role in my recent work
on nonequilibrium statistical mechanics. This is another story
for which I refer to my paper 
\emph{From fluctuations in hydrodynamics to nonequilibrium thermodynamics}
in this same issue of the Progress of Theoretical Physics Supplement.

\section*{Acknowledgements}
This special seminar within the activities of the 2009 YKIS workshop on 
nonequilibrium statistical mechanics was prompted by Shin-ichi Sasa.
To him and all the organizers I wish to express my gratitude for the
invitation.

%

\end{document}